\documentclass[prb,nofootinbib,twocolumn,superscriptaddress]{revtex4} 

\usepackage{graphicx}
\usepackage{dcolumn}
\usepackage{bm}
\usepackage{threeparttable}
\usepackage{times}
\usepackage{mathptmx}
\usepackage{lscape}
\usepackage{natbib}
\usepackage{amsmath}
\usepackage{amssymb}
\usepackage{braket}
\usepackage{comment}
\usepackage{color}

\def\degree{\kern-.2em\r{}\kern-.3em}

\begin{document}


\title{ Universal Characterization of Hierarchical Ordering Tendency\\ in High-Entropy Alloys from Configurational Geometry   }

\author{Koretaka Yuge}
\affiliation{
Department of Materials Science and Engineering,  Kyoto University, Sakyo, Kyoto 606-8501, Japan\\
}%

\begin{abstract}
{ Microscopic structures for fcc-based quaternary high-entropy alloys (HEA) in thermodynamically equilibrium state is examined based on first-principles (FP) calculation combined with our recently-developed theoretical approach. We find that (i) hierarchical ordering tendency for whole quaternary, and ternary and binary subsystems for the present five HEAs cannot be reasonably characterized by conventional Goldschmidt atomic radius or by those obtained from unary system for FP calculation, but can be systematically characterized by atomic radius from specially fluctuated configuration. (ii) ordering tendency for whole- and sub-systems can be simultaneously treated with individual definition of atomic radius ratio, and that linear correlation between ordering tendency and atomic radius ratio naturally decreases with decrease of number of constituents for whole (or sub-) system due to the counterbalance breaking of like- and unlike-atom neighboring pair. We also find that by introducing appropriate normalization, ordering tendency for systems with different number of constituents (here, quaternary and ternary) can also be simultaneously treated by the atomic radius ratio. 
  }
\end{abstract}


\maketitle

\section{Introduction}
Short-range ordering (SRO) tendency in alloys has been amply investigated by experimental as well as theoretical approaches, especially for binary disordered alloys, since it has significant correlation with e.g., their ground-state structures and mechanical and functional properties. 
It has been firstly considered that for binary alloys, constituent atomic radius ratio can characterize the SRO tendency, due mainly to the intuition that ordering tendency, i.e., unlike-atom pair preference, can be naturally enhanced by reducing strain energy when the ratio aparts from one. 
However, following theoretical studies quantitatively find that such intuition easily fails, since SRO tendency should be determined by competition between chemical ordering and geometric (i.e., atomic radius ratio) effects where the former typically plays central roles.\cite{sro-j,chem1,chem2} 
For multicomponent (i.e., number of constituents not less than three) alloys, the situation becomes more complicated: Since counterbalance of unlike- and like-atom pair for constituent subsystems can break, we can easily expect that such hierarchical ordering tendency for multicomponent alloys is further difficult to be characterized only from geometric information. 

Despite these facts, we recently find\cite{em-sro} through first-principles calculation combined with our recently-developed theoretical approach, that SRO tendency for 27 fcc-based binary alloys, as well as hierarchical SRO for 7 fcc-based ternary alloys, can be respectively well-characterized by atomic radius ratio obtained from specially-fluctuated structure constructed by configurational geometry independent of energy and of temperature. The results strongly indicate the significant role of information about configurational geometry on ordering tendencies in multicomponent alloys. 
The present study extends our previous work, to confine whether or not (i) such geometric information can still characterize higher order hierarchical SRO with increase of number of components to quaternary alloys, and (ii) when (i) is achieved, SRO for different number of components can be simultaneously treated. The details are discussed below. 

\section{Methodology}
\subsection{Pair Probability}
In order to quantitatively, completely describe any atomic configuration on given lattice, generalized Ising model\cite{ce} (GIM) is first employed to provide complete orthonormal basis functions. 
For the present A-B-C-D quaternary systems, such basis on a single lattice point is given by 
\begin{eqnarray}
\label{eq:basis}
&&\phi_0 = 1,\quad \phi_1 =\frac{2}{\sqrt{10}}\sigma_i \nonumber \\
&&\phi_2 = -\frac{5}{3} + \frac{2}{3}\sigma_{i}^{2},\quad \phi_{3} = -\frac{17\sqrt{10}}{30}\sigma_{i} + \frac{\sqrt{10}}{6}\sigma_{i}^{3}
\end{eqnarray}
with definition of spin variable $\sigma_{i}$ to specify occupation at lattice point $i$ by element A, B, C or D respectively leading to $\sigma_{i}=\left\{2,1,-1,-2\right\}$.
Higher-order multisite correlations $\Phi$ (e.g., $\Phi_{pq}=\phi_{p}\phi_{q}$)can be obtained by taking average of products of the above basis functions over all lattice points. 
Based on the GIM orthonormal basis $\left\{\Phi\right\}$, we recently derive a formulation to characterize temperature dependence of pair probability $y_{IJ}$ for disordered states with any number of components, given by\cite{ps-multi} 
\begin{widetext}
\begin{eqnarray}
\label{eq:y}
\Braket{y_{IJ}}\left(T\right) \simeq \Braket{y_{IJ}}_1 - \sqrt{\frac{\pi}{2}} \frac{\Braket{y_{IJ}}_2}{k_{\textrm{B}}T}\sum_M \Braket{U|\Phi_M}   \left( \Braket{\Phi_M}_{y_{IJ}}^{\left(+\right)} - \Braket{\Phi_M}_1 \right).
\end{eqnarray}
\end{widetext}
Here, $\Braket{\quad}_{1}$ and $\Braket{\quad}_{2}$ respectively denotes taking linear average and standard deviation for configurational density of states (CDOS) along corresponding coordination, \textit{before} applying many-body interaction to the system. $\Braket{\quad}^{\left(+\right)}_{Y_{IJ}}$  denotes taking partial average over configuration $\vec{\sigma}$satisfying $y_{IJ}\left(\vec{\sigma}\right) \ge \Braket{y_{IJ}}_{1}$, and $\Braket{\quad|\quad}$ represents inner product on configuration space. 
The definition of pair probability is explicitly given by the following relationships:
\begin{eqnarray}
\sum_{J} y_{IJ} = c_I,
\end{eqnarray}
where $c_{I}$ denotes composition for component $I$.
Equation~\eqref{eq:y} certainly indicate that $f$-dimensional configuration space, when structure of two atomic configurations are respectively given by
\begin{eqnarray}
\label{eq:psy}
&&\left\{ \Braket{\Phi_1}_{y_{IJ}}^{\left(+\right)},\cdots, \Braket{\Phi_f}_{y_{IJ}}^{\left(+\right)} \right\} \quad \left(\mathrm{str. 1}\right)\nonumber \\
&&\left\{\Braket{\Phi_1}_1,\cdots,\Braket{\Phi_f}_1\right\} \quad \left(\mathrm{str. 2}\right),
\end{eqnarray}
Eq.~\eqref{eq:y} can be rewritten by potential energy for str. 1 ($U_{IJ}$) and for str. 2 ($\left(U_{0}\right)$):
\begin{eqnarray}
\Braket{y_{IJ}}\left(T\right) \simeq \Braket{y_{IJ}}_1 - \sqrt{\frac{\pi}{2}} \frac{\Braket{y_{IJ}}_2}{k_{\textrm{B}}T}\left(U_{IJ} - U_{0}\right).
\end{eqnarray}
Str. 1 and Str. 2 are respectively called projection state\cite{em2} (PS) and special quasirandom structure\cite{sqs} (SQS). 
Here, the important point is that from Eq.~\eqref{eq:psy}, structures of PS and SQS can be known \textit{a priori} without any information about energy or temperature:
They are then constructed by performing Monte Calro (MC) simulation to uniformly sampling possible configurations on fcc quaternary-equiatomic composition with supercell of 480-atom (i.e., $4\times5\times6$ expansion of conventional cubic unit cell with 4-atoms), to find optimal structures satisfying (i.e., minimizing Euclidean distance) Eq.~\eqref{eq:psy}. 
We here consider configuration space consisting of up to 6NN pair, and all triplets and quartets consisting of up to 4NN pairs that can typically well characterize ordering tendency for fcc-based binary alloys.\cite{em-sro}
To practically construct structures of PS and SQS, we employ the following relationships between like- and unlike-pair probability and GIM basis for quaternary system:\cite{ps-multi}
\begin{widetext}
\begin{eqnarray}
\label{eq:4y}
y_{AA} &=& \frac{1}{2}c_a+\frac{1}{10}\Phi_{11}+\frac{\sqrt{10}}{20}\Phi_{12}+\frac{1}{10}\Phi_{13}+\frac{1}{16}\Phi_{22}+\frac{\sqrt{10}}{40}\Phi_{23}+\frac{1}{40}\Phi_{33}-\frac{1}{16} \nonumber \\
y_{AB} &=& \frac{1}{4}c_a+\frac{1}{4}c_b+\frac{1}{20}\Phi_{11}-\frac{\sqrt{10}}{80}\Phi_{12}-\frac{3}{40}\Phi_{13}-\frac{1}{16}\Phi_{22}-\frac{3\sqrt{10}}{80}\Phi_{23}-\frac{1}{20}\Phi_{33}-\frac{1}{16}. \nonumber \\
\end{eqnarray}
\end{widetext}
Constructed PSs and SQS are applied to density functional theory (DFT) calculation to obtain total energy for five quaternary equiatomic alloys of CrMnFeCo, CrMnFeNi, CrMnCoNi, CrFeCoNi and MnFeCoNi, performed by the VASP\cite{vasp} code using projector-augmented wave method,\cite{paw} with the exchange-correlation functional treated within the generalized-gradient approximation of Perdew-Burke- Ernzerhof (GGA-PBE).\cite{pbe} The plane wave cutoff of 360 eV is used. Structural optimization is performed until the residual forces less than 0.005 eV/\AA. 

\subsection{Description of Hierarchical Ordering Tendency}
Since for multicomponent alloys, counterbalance between like- and unlike-atom pair ordering can be broken for subsystems, hierarchical ordering tendency naturally appears for whole quaternary system, and ternary and binary subsystems. To systematically and simultaneously treat such ordering tendency, we here introduce 3 definitions to measure the tendency for each layer (slight modification and extension of our previously defined one, enabling to directly compare ordering tendency in different subsystems):
\begin{eqnarray}
\label{eq:a}
\alpha_{4} &=& \frac{1}{4}\sum_{J} U_{JJ} - \frac{1}{6} \sum_{I\neq J} U_{IJ} \nonumber \\
\alpha_{3}^{\left(IJK\right)} &=& \frac{1}{3}\sum_{p\in\left\{I,J,K\right\}} U_{pp} - \frac{1}{3}\sum_{\substack{p, q \in \left\{I,J,K\right\} \\ p\neq q}} U_{pq} \nonumber \\
\alpha_{2}^{\left(IJ\right)} &=& \frac{1}{2} \left(U_{II} + U_{JJ}\right) - U_{IJ},
\end{eqnarray}
where for simplicity, here and hereinafter we describe energy and structure measured from SQS energy and its correlation function. 
From the definitions in Eq.~\eqref{eq:a}, we can see that positive (negative) sign of $\alpha_{4}$, $\alpha_{3}$ and $\alpha_{2}$ respectively corresponds to preference (disfavor) of unlike-atom pair(s) w.r.t. the rest like-atom pairs in quaternary system, and ternary and binary subsystems, and ideally random states should satisfy $\alpha_{4}=\alpha_{3}=\alpha_{2}=0$. 

To characterize the above three definition of ordering tendencies, we correspondingly prepare three types of atomic radius ratio for binary subsystem:
\begin{eqnarray}
R_{g}^{\left(JK\right)} &=& \left[ \frac{R_{g}^{\left(J\right)} }{R_{g}^{\left(K\right)}}\right]_{1} \nonumber \\
R_{u}^{\left(JK\right)} &=& \left[ \frac{R_{u}^{\left(J\right)} }{R_{u}^{\left(K\right)}}\right]_{1} \nonumber \\
R_{PS}^{\left(JK\right)} &=& \left[ \frac{R_{PS}^{\left(J\right)} }{R_{PS}^{\left(K\right)}}\right]_{1}, 
\end{eqnarray}
where $\left[\quad\right]_{1}$ denotes that internal numerator and denominator can be reversed so that resultant value of fraction should always take greater or equal to 1.
$R_{g}$, $R_{u}$ and $R_{PS}$ respectively denotes atomic radius obtained from Goldschmidt atomic radius,\cite{gold} unary system by FP calculation, and by PS with solving algebraic equations for multiple projection state energies.\cite{sro-multi}

\section{Results and Discussions}
\begin{figure}[h]
\begin{center}
\includegraphics[width=1.03\linewidth]{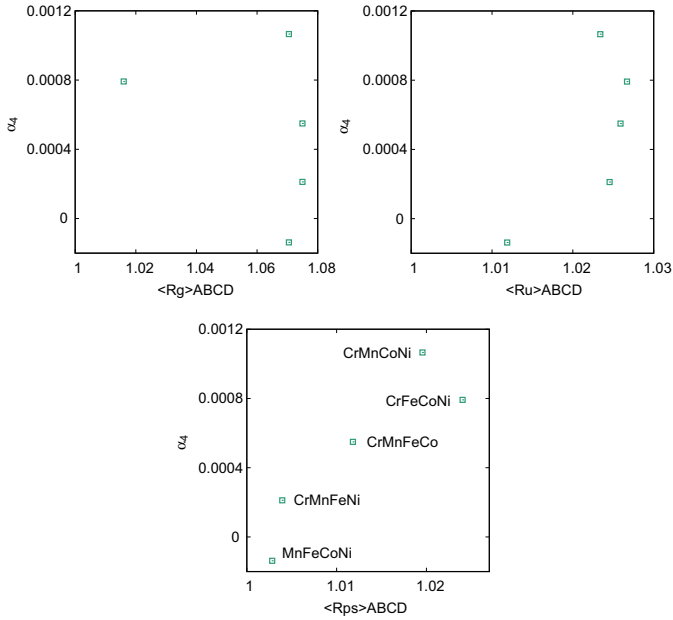}
\caption{Quaternary ordering tendency $\alpha_{4}$ in terms of ratio of constituent atomic radius averaged over possible pairs in 4 components.}
\label{fig:a4}
\end{center}
\end{figure}
\begin{figure}[h]
\begin{center}
\includegraphics[width=1.03\linewidth]
{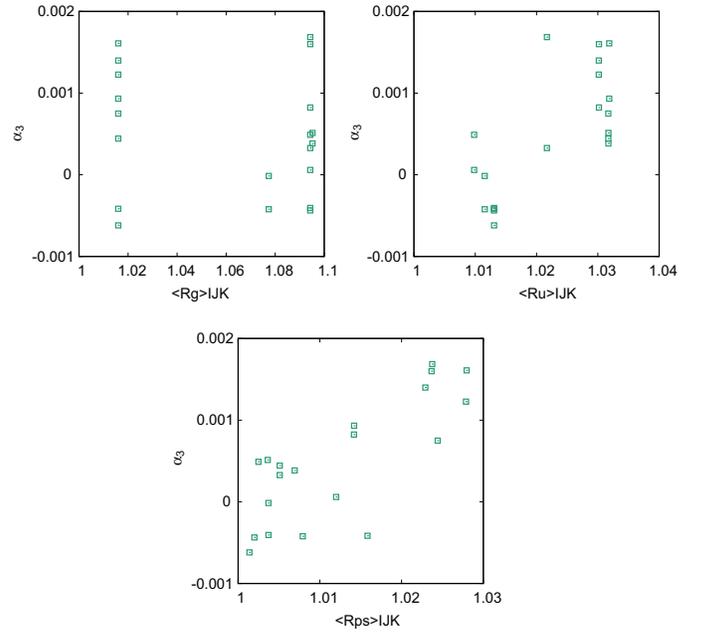}
\caption{Ordering tendency $\alpha_{3}$ for ternary subsystems in terms of ratio of constituent atomic radius averaged over possible pairs in 3 components.}
\label{fig:a3}
\end{center}
\end{figure}
\begin{figure}[h]
\begin{center}
\includegraphics[width=1.03\linewidth]
{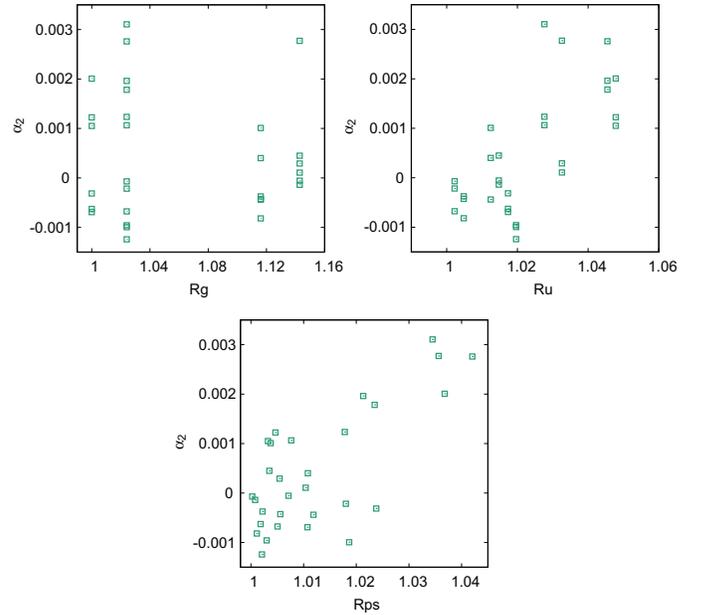}
\caption{Ordering tendency $\alpha_{2}$ for binary subsystems in terms of ratio of constituent atomic radius.}
\label{fig:a2}
\end{center}
\end{figure}
Figures~\ref{fig:a4}-\ref{fig:a2} shows the resultant value of $\alpha_{4}$, $\alpha_{3}$ and $\alpha_{2}$ for the five quaternary HEAs in terms of the three definition of atomic radius ratio. 
Here, the brackets denote taking linear average of atomic radius ratio for individual whole (or sub) systems
\begin{eqnarray}
\Braket{R}_{ABCD} &=& \frac{1}{6}\sum_{\substack{I,J\in\left\{A,B,C,D\right\}\\ I\neq J  }} R^{\left(IJ\right)}  \nonumber \\
\Braket{R}_{IJK} &=& \frac{1}{3}\sum_{\substack{p,q\in\left\{I,J,K\right\} \\ p\neq q }  } R^{\left(pq\right)}.
\end{eqnarray}
We can clearly see that (i) conventional Goldschmidt atomic radius cannot capture ordering tendency for whole or any subsystems, (ii) $R$ from unary system has weak correlation with SRO for whole and subsystems due mainly to the poor information about atomic radius in preferring ordering configuration and thus to the poor variety of resultant atomic radius, and (iii) $R$ from projection state, i.e., explicitly including covariance fluctuation between structural degree of freedoms in configurational geometry,\cite{spe} can universally well-characterize ordering tendency for whole as well as subsystems. 
\begin{figure}[h]
\begin{center}
\includegraphics[width=0.6\linewidth]{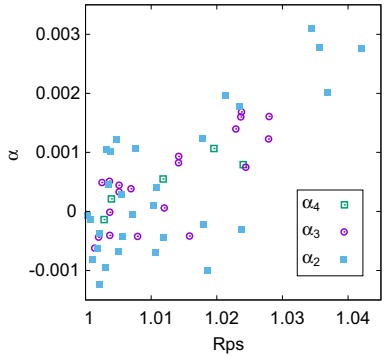}
\caption{Merged hierarchical ordering tendency $\alpha_{4}$, $\alpha_{3}$ and $\alpha_{2}$ in terms of ratio of constituent atomic radius in respective number of components obtained from special microscopic states. }
\label{fig:merge}
\end{center}
\end{figure}
Therefore, we then marginally plot $\alpha_{4}$, $\alpha_{3}$ and $\alpha_{2}$ with atomic radius ratio $R_{PS}$, shown in Fig.~\ref{fig:merge}. We can clearly see that ordering tendency for whole and subsystems in quaternary alloys can be simultaneously characterized by $R_{PS}$ and their averaged value for corresponding constituent systems. Decrease of linear correlation with decrease of constituents (i.e., from quaternary to ternary subsystem and to binary subsystem) can be naturally interpreted by enhance of counterbalance breaking between like- and unlike-atomic pair probability. 

\begin{figure}[h]
\begin{center}
\includegraphics[width=0.6\linewidth]{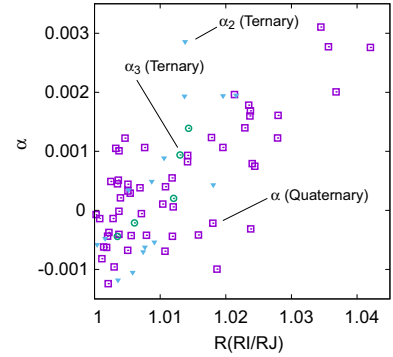}
\caption{Merged hierarchical ordering tendency for quaternary and pure-ternary system. }
\label{fig:mgqt}
\end{center}
\end{figure}

Since complicated hierarchical ordering tendency for quaternary alloys can be reasonably characterized by the same definition of atomic radius ratio, we further confirm whether or not such features with systems for \textit{different}  number of components (here, the present quaternary and our previous study of ternary alloys\cite{sro-multi}). Here, the problem is that due to the difference in number of components, values of $\alpha$ for quaternary and ternary systems cannot be straightforwardly compared due mainly to the essential differences in composition as well as effective changes in ordering parameter from ideally random state coming from difference in variance of CDOS. To simultaneously overcome these problems, we naturally introduce normalization for ternary ordering tendency, given by
\begin{eqnarray}
\alpha'_{\textrm{ternary}} = \left( \frac{\Braket{Y_{IJ}}_{1}^{\left( 3 \right)}}{ \Braket{Y_{IJ}}_{1}^{\left( 4 \right)}} \right)\cdot \left( \frac{\Braket{Y_{IJ}}_{2}^{\left( 4 \right)}}{ \Braket{Y_{IJ}}_{2}^{\left( 3 \right)}} \right)\cdot \alpha_{\textrm{ternary}},
\end{eqnarray}
where $\alpha_{\textrm{ternary}}$ is the same definition for present $\alpha$s in quaternary systems, and superscript $\left( 4 \right)$ and $\left( 3 \right)$ respectively denotes average (or standard deviation) for quaternary and ternary system. With this definition, we add results in our previous study for five ternary equiatomic alloys of CrCoNi, CrFeNi, CoNiMn, FeNiMn and CrNiMn where the constituent elements are the same as the present quaternary alloys, shown in Fig.~\ref{fig:mgqt} .
We can see that with the present normalization to ternary systems, ordering tendency for quaternary, its ternary and binary subsystems, and ternary, and its binary subsystems can be qualitatively well-characterized throughout atomic radius ratio obtained from special microscopic state, PS. 
These findings strongly indicate that configurational geometry can universally bridge complicated hierarchical ordering tendency for multicomponent alloys.

\section{Conclusions}
We theoretically examine hierarchical ordering tendency of multicomponent HEAs in thermodynamically equilibrium state. We find that ordering tendency for whole system, its derivative subsystems as well as whole systems with different number of constituents, can be universally well-characterized by atomic radius ratio obtained from special microscopic states reflecting covariance fluctuation between structural degree of freedoms, These findings strongly indicate significant role of configurational geometry \textit{before} applying many-body interaction to the system to determine characteristic ordering tendencies for multicomponent alloys.

\section{Acknowledgement}
This work was supported by Grant-in-Aids for Scientific Research on Innovative Areas on High Entropy Alloys through the grant number JP18H05453 and a Grant-in-Aid for Scientific Research (16K06704) from the MEXT of Japan, Research Grant from Hitachi Metals$\cdot$Materials Science Foundation, and Advanced Low Carbon Technology Research and Development Program of the Japan Science and Technology Agency (JST).


\begin{thebibliography}{9}
\bibitem{sro-j} J. Cryst. Soc. Jpn. \textbf{12}, 186 (1970).
\bibitem{chem1} R. V. Chepulskii, J. Phys.: Condens. Matter \textbf{10}, 1505 (1998).
\bibitem{chem2} R. V. Chepulskii and V. N. Bugaev, J. Phys.: Condens. Matter \textbf{10}, 7309 (1998).
\bibitem{em-sro} K. Yuge, J. Phys. Soc. Jpn.  \textbf{87}, 044804 (2018).
\bibitem{ce} J.M. Sanchez, F. Ducastelle, and D. Gratias, Physica A \textbf{128}, 334 (1984).
\bibitem{ps-multi} K. Yuge and S. Ohta,  J. Phys. Soc. Jpn. \textbf{88}, 044803 (2019). 
\bibitem{em2} K. Yuge, J. Phys. Soc. Jpn. \textbf{85}, 024802  (2016).
\bibitem{sqs} S.-H. Wei, L. G. Ferreira, J. E. Bernard, and A. Zunger, Phys. Rev. B \textbf{42}, 9622 (1990).
\bibitem{vasp} G. Kresse and J. Hafner, Phys. Rev. B {\textbf{47}}, R558 (1993).
\bibitem{paw} G. Kresse and D. Joubert, Phys. Rev. B {\textbf{59}}, 1758 (1999). 
\bibitem{pbe} J.P. Perdew, K. Burke, and M. Ernzerhof, Phys. Rev. Lett. {\textbf{77}}, 3865 (1996).
\bibitem{gold} V. M. Goldschmidt, Z. Phys. Chem. \textbf{133}, 397 (1928).
\bibitem{spe} K. Yuge,  J. Phys. Soc. Jpn. \textbf{87}, 104802-1-6 (2018). 
\bibitem{sro-multi} K. Yuge and S. Ohta,  J. Phys. Soc. Jpn. \textbf{88}, 054803 (2019).
\end{thebibliography}
\end{document}